\documentclass{article}
\usepackage{amsmath,upref}
\usepackage{amssymb,amsbsy}
\usepackage{amsfonts}

\title{\textbf{Constrained systems and the Clairaut equation}}
\author{\textbf{Steven Duplij}\\Institut f\"ur Theoretische Physik,
Universit\"at zu K\"oln,\\ Z\"ulpicher Str. 77, 50937 K\"oln,
Germany\thanks{\emph{On leave of absence from}: Theory Group,
Nuclear Physics Laboratory, V. N. Karazin Kharkov National
University, Svoboda Sq. 4, Kharkov 61077, Ukraine,
\texttt{sduplij@gmail.com},
http://webusers.physics.umn.edu/\~{}duplij.}\\
E-mail: duplij@thp.uni-koeln.de}

\date{April 16, 2008 \\\small (Revised September 12, 2009)}

\begin{document}

\maketitle

\begin{abstract}
An extension of the Legendre transform to non-convex functions with
vanishing Hessian as a mix of envelope and general solutions of the
Clairaut equation is proposed.  Applying this to systems with  constraints,
the procedure of finding  a Hamiltonian for a degenerate Lagrangian is
just that of solving a corresponding Clairaut equation with a
subsequent application of the proposed Legendre-Clairaut
transformation. In this way the unconstrained version of Hamiltonian
equations is obtained. The Legendre-Clairaut transformation
presented is involutive. We demonstrate the origin of the Dirac
primary constraints, along with their explicit form, and this is
done without using the Lagrange multiplier method.

\end{abstract}

\section{Introduction}

Modern field theories are in fact degenerate dynamical systems whose key
feature is the presence of constraints \cite{sud/muk,sundermeyer}. The most
common way to deal with such systems is to use the Dirac approach \cite%
{dirac} based on extending a phase space and constructing the so-called
total Hamiltonian. In spite of its general success, e.g. in describing
systems with gauge symmetries and gravity \cite{reg/tei,git/tyu}, the Dirac
approach has limited applicability and some inner problems \cite{pon3}.
Therefore, it is worthwhile to reconsider several basic ideas of the
Hamiltonian formalism per se starting from the Legendre transformation
treated as a solution of the Clairaut equation also in the singular case.
The Dirac approach is based on the following idea: in finding a Hamiltonian,
to use the standard definition of momenta, then perform the Legendre
transformation and add Lagrange multipliers, subsequently to be removed by
imposing some relations between constraints \cite{dirac}.

We revisit the procedure of finding a Hamiltonian for both regular and
singular cases. This procedure is reduced to that of solving the Clairaut
partial differential equation. In the case of regular systems a Hamiltonian
corresponds to its envelope solution \cite{arnold}, while adding the general
solution of Clairaut equation leads to the total Hamiltonian of singular
systems, and arbitrary constants correspond to the Lagrange multipliers
within the Dirac approach \cite{dirac}. Such solutions exist for smooth
Lagrangian functions, while the standard Legendre transformation is
applicable in the regular case only. To solve the Clairaut equation in the
singular case, we introduce a mixed envelope solution, which is an envelope
solution in ``regular'' variables and a general solution in ``nonregular''
variables. We use the coordinate language which is convenient for making the
basic idea transparent as well as for further applications.

\section{ \label{sec-lt}The Legendre transform and the Clairaut equation}

We recall the standard Legendre transform definition \cite{arnold}. The
Legendre transform of a convex function\footnote{%
We use vector notation \cite{arnold} in which $\mathbf{x}\in \overset{n}{%
\overbrace{\mathbb{R}\times \mathbb{R}\times \ldots \mathbb{R}}}=\mathbb{R}%
^{n}$, and $\mathbf{x}\cdot \mathbf{y}$ is the scalar product. Functions and
variables are denoted by capitals and lowercase letters, respectively. A
scalar differentiable function $F:\mathbb{R}^{n}\rightarrow \mathbb{R}$ is
denoted by $F\left( \mathbf{x}\right) $. A vector $\partial F\diagup
\partial \mathbf{x}$ denotes a gradient of $F$ whose entries are just
partial derivatives $\partial F\diagup \partial x^{i}$, $i=1,\ldots \left[ i%
\right] =n$, and for $\partial ^{2}F\diagup \partial x^{i}\partial x^{j}$ we
sometimes use the notation $\partial ^{2}F\diagup \partial \mathbf{x}^{2}$.
Also, $\partial \mathbf{V}\diagup \partial \mathbf{x}$ denotes the
divergence of a vector function $\mathbf{V}:\mathbb{R}^{n}\rightarrow
\mathbb{R}^{n}$ in a similar manner.} $F:\mathbb{R}^{n}\rightarrow \mathbb{R}
$ is a map $\mathfrak{Leg}:F\longmapsto G$, where $G\left( \mathbf{p}\right)
:\mathbb{R}^{n\ast }\rightarrow \mathbb{R}$ is another convex function (in
the dual space) such that $G\left( \mathbf{p}\right) =\max {}_{\mathbf{x}}%
\tilde{G}\left( \mathbf{p},\mathbf{x}\right) $, and $\tilde{G}\left( \mathbf{%
p},\mathbf{x}\right) \overset{def}{=}\mathbf{p}\cdot \mathbf{x}-F\left(
\mathbf{x}\right) $. The maximum is attained, when%
\begin{equation}
\dfrac{\partial \tilde{G}\left( \mathbf{p},\mathbf{x}\right) }{\partial
\mathbf{x}}=\mathbf{p}-\dfrac{\partial F\left( \mathbf{x}\right) }{\partial
\mathbf{x}}=0,  \label{p}
\end{equation}%
which, for a given $\mathbf{p}$, can determine $\mathbf{x}$ unambiguously,
but not vice versa. The convexity implies that the Hessian $H_{F}\left(
\mathbf{x}\right) \overset{def}{=}\det \left\Vert \tfrac{\partial
^{2}F\left( \mathbf{x}\right) }{\partial \mathbf{x}^{2}}\right\Vert $ is
positive, which ensures that (globally) there is only one critical point,
the maximum \cite{arnold}.

Now we reformulate the Legendre transform in terms of the differential
equation for $G\left( \mathbf{p}\right) $ as above. Let us suppose that
\textbf{(}\ref{p}\textbf{)} has a solution $\mathbf{x}=\mathbf{X}\left(
\mathbf{p}\right) $ with $\mathbf{X}:\mathbb{R}^{n}\rightarrow \mathbb{R}^{n}
$, then
\begin{equation}
G\left( \mathbf{p}\right) \overset{def}{=}\tilde{G}\left( \mathbf{p},\mathbf{%
X}\left( \mathbf{p}\right) \right) =\mathbf{p\cdot X}\left( \mathbf{p}%
\right) -F\left( \mathbf{X}\left( \mathbf{p}\right) \right) .  \label{g0}
\end{equation}%
Since $\tfrac{\partial ^{2}G\left( \mathbf{p}\right) }{\partial \mathbf{p}%
^{2}}\overset{\left( \ref{g0}\right) }{=}\left( \tfrac{\partial \mathbf{X}%
\left( \mathbf{p}\right) }{\partial \mathbf{p}}\right) ^{2}\tfrac{\partial
^{2}F\left( \mathbf{x}\right) }{\partial \mathbf{x}^{2}}$, the functions $F$
and $G$ belong to the same convexity class and in the case of convex
functions one has $rank\left\Vert \tfrac{\partial ^{2}G\left( \mathbf{p}%
\right) }{\partial \mathbf{p}^{2}}\right\Vert =rank\left\Vert \tfrac{%
\partial ^{2}F\left( \mathbf{x}\right) }{\partial \mathbf{x}^{2}}\right\Vert
$. Next we differentiate (\ref{g0}) and get $\tfrac{\partial G\left( \mathbf{%
p}\right) }{\partial \mathbf{p}}\overset{(\ref{p})}{=}\mathbf{X}\left(
\mathbf{p}\right) $, which allows us to exclude $\mathbf{X}\left( \mathbf{p}%
\right) $ from (\ref{g0}) and obtain%
\begin{equation}
G^{Cl}\left( \mathbf{p}\right) =\mathbf{p}\cdot \dfrac{\partial G^{Cl}\left(
\mathbf{p}\right) }{\partial \mathbf{p}}-F\left( \dfrac{\partial
G^{Cl}\left( \mathbf{p}\right) }{\partial \mathbf{p}}\right) ,  \label{g}
\end{equation}%
which is just the Clairaut equation \cite{arnold1}. Obviously, (\ref{g}) has
solutions even in cases, where (\ref{p}) cannot be resolved in $\mathbf{x}$,
and therefore the standard Legendre transform (\ref{g0}) does not exist
(thus we add the superscript $Cl$). We call a map $\mathfrak{Leg}%
^{Cl}:F\longmapsto G^{Cl}$ defined by (\ref{g}) a (generalized)
Legendre-Clairaut transform with $F\left( \mathbf{x}\right) $ being any
(smooth) function (not necessarily convex). For a convex function $F\left(
\mathbf{x}\right) $, if $G\left( \mathbf{p}\right) =Leg_{F}\left( \mathbf{p}%
\right) $ is its standard Legendre transform, then $G\left( \mathbf{p}%
\right) $ satisfies (\ref{g}), hence $G\left( \mathbf{p}\right) $ coincides
with $G^{Cl}\left( \mathbf{p}\right) =Leg_{F}^{Cl}\left( \mathbf{p}\right) $%
, the Legendre-Clairaut transform of $F\left( \mathbf{x}\right) $. Now we
demonstrate the converse statement. Let us write a general solution of (\ref%
{g}) as%
\begin{equation}
\tilde{G}_{gen}^{Cl}\left( \mathbf{p},\mathbf{c}\right) =\mathbf{p}\cdot
\mathbf{c}-F\left( \mathbf{c}\right) ,  \label{gen}
\end{equation}%
where $\mathbf{c}\in \mathbb{R}^{N}$. The envelope solution can be obtained
from the extremum condition $\tfrac{\partial \tilde{G}_{gen}^{Cl}\left(
\mathbf{p},\mathbf{c}\right) }{\partial \mathbf{c}}=\mathbf{p}-\tfrac{%
\partial F\left( \mathbf{c}\right) }{\partial \mathbf{c}}=0$, which
coincides with the condition (\ref{p}) (this $\tilde{G}_{gen}^{Cl}$ actually
coincides with $\tilde{G}$ above). If $H_{F}\left( \mathbf{c}\right) >0$ ($F$
is convex), the extremum condition can be solved by $\mathbf{c}=\mathbf{X}%
\left( \mathbf{p}\right) $. Then the envelope solution of (\ref{g}) is%
\begin{equation}
G_{env}^{Cl}\left( \mathbf{p}\right) \overset{def}{=}\tilde{G}%
_{gen}^{Cl}\left( \mathbf{p},\mathbf{X}\left( \mathbf{p}\right) \right) =%
\mathbf{p\cdot X}\left( \mathbf{p}\right) -F\left( \mathbf{X}\left( \mathbf{p%
}\right) \right) \overset{\left( \ref{g0}\right) }{=}G\left( \mathbf{p}%
\right) .  \label{env}
\end{equation}%
This means that, given a convex function $F\left( \mathbf{x}\right) $, if $%
G^{Cl}\left( \mathbf{p}\right) =Leg_{F}^{Cl}\left( \mathbf{p}\right) $, then
$G^{Cl}\left( \mathbf{p}\right) =Leg_{F}\left( \mathbf{p}\right) =G\left(
\mathbf{p}\right) $. Also, only in this (convex) case both mappings $%
\mathfrak{Leg}^{Cl}$ and $\mathfrak{Leg}$ are involutive $\mathfrak{Leg}%
^{Cl}\circ \mathfrak{Leg}^{Cl}=id$, $\mathfrak{Leg}\circ \mathfrak{Leg}=id$.
Thus, \textit{the standard Legendre transforms in the class of convex
functions are in 1-1 correspondence with the envelope solutions of the
Clairaut equation} (\ref{g}). This provides an exposition of the ordinary
theory \cite{arnold,arnold1} in a special way, which is convenient for our
subsequent purposes that involve more general classes of functions related
to the Hamiltonian structure of constrained systems \cite{dirac,git/tyu}.

Indeed, let us consider the case of a non-convex function $F\left( \mathbf{x}%
\right) $, when its Hessian $H_{F}\left( \mathbf{x}\right) $ vanishes.
The standard Legendre trick does not work, because (\ref{p}) cannot be
solved for $\mathbf{x}$ in this case \cite{arnold}. On the other hand, the
Clairaut equation (\ref{g}) assumes nothing but smoothness about $F\left(
\mathbf{x}\right) $. Therefore, we can forget the condition of its obtaining
and start from the Clairaut equation itself, then try to find the solutions.
In this way, we can extend the Legendre-Clairaut transform to the degenerate
case $H_{F}\left( \mathbf{x}\right) =0$. Let $rank\left\Vert \tfrac{\partial
^{2}F\left( \mathbf{x}\right) }{\partial \mathbf{x}^{2}}\right\Vert =k<n$
and $k$ is constant on the domain of $F\left( \mathbf{x}\right) $. Without
loss of generality, we can assume the indices are rearranged in such a way
that a non-singular minor of rank $k$ is in the upper left-hand corner.
Then, we express the index $i$ as a pair $i=\left( i_{1},i_{2}\right) $, $%
i_{1}=1,\ldots k$, $i_{2}=k+1,\ldots n$, and correspondingly any vector
variable is presented as $\mathbf{x}=\left( \mathbf{x}_{1},\mathbf{x}%
_{2}\right) $, $\mathbf{x}_{1}\in \mathbb{R}^{k}$, $\mathbf{x}_{2}\in
\mathbb{R}^{n-k}$. We call the first and the second entry of $\mathbf{x}$ as
regular and nonregular, respectively. In this notation, the Clairaut
equation (\ref{g}) becomes%
\begin{eqnarray}
G^{Cl}\left( \mathbf{p}_{1},\mathbf{p}_{2}\right)  &=&\mathbf{p}_{1}\cdot
\dfrac{\partial G^{Cl}\left( \mathbf{p}_{1},\mathbf{p}_{2}\right) }{\partial
\mathbf{p}_{1}}+\mathbf{p}_{2}\cdot \dfrac{\partial G^{Cl}\left( \mathbf{p}%
_{1},\mathbf{p}_{2}\right) }{\partial \mathbf{p}_{2}}  \notag \\
&&-F\left( \dfrac{\partial G^{Cl}\left( \mathbf{p}_{1},\mathbf{p}_{2}\right)
}{\partial \mathbf{p}_{1}},\dfrac{\partial G^{Cl}\left( \mathbf{p}_{1},%
\mathbf{p}_{2}\right) }{\partial \mathbf{p}_{2}}\right) .  \label{gl2}
\end{eqnarray}%
\ By analogy with (\ref{gen}), the general solution is%
\begin{equation}
\tilde{G}_{gen}^{Cl}\left( \mathbf{p}_{1},\mathbf{p}_{2},\mathbf{c}_{1},%
\mathbf{c}_{2}\right) =\mathbf{p}_{1}\cdot \mathbf{c}_{1}+\mathbf{p}%
_{2}\cdot \mathbf{c}_{2}-F\left( \mathbf{c}_{1},\mathbf{c}_{2}\right) ,
\label{c12}
\end{equation}%
where $\mathbf{c}_{1}\in \mathbb{R}^{k}$, $\mathbf{c}_{2}\in \mathbb{R}^{n-k}
$. Our intention now is to search for the envelope solution (\ref{env}) in
regular variables only. But it is still the general solution (\ref{gen})
with arbitrary $\mathbf{c}_{2}$ with respect to non-regular variables. We
call such a solution \textit{a mixed envelope solution of the Clairaut
equation}. Differentiate (\ref{c12}) in $\mathbf{c}_{1}$ to get%
\begin{equation}
\dfrac{\partial \tilde{G}_{gen}^{Cl}\left( \mathbf{p}_{1},\mathbf{p}_{2},%
\mathbf{c}_{1},\mathbf{c}_{2}\right) }{\partial \mathbf{c}_{1}}=\mathbf{p}%
_{1}-\dfrac{\partial F\left( \mathbf{c}_{1},\mathbf{c}_{2}\right) }{\partial
\mathbf{c}_{1}}=0  \label{d12}
\end{equation}%
Since the sub-Hessian $H_{F}^{\left( 1\right) }\left( \mathbf{x}\right)
\overset{def}{=}\det \left\Vert \tfrac{\partial ^{2}F\left( \mathbf{x}_{1},%
\mathbf{x}_{2}\right) }{\partial \mathbf{x}_{1}^{2}}\right\Vert $ does not
vanish, we can resolve (\ref{d12}) with respect to $\mathbf{c}_{1}$ and
obtain $\mathbf{c}_{1}=\mathbf{X}\left( \mathbf{p}_{1},\mathbf{c}_{2}\right)
$. The subsequent substitution to (\ref{c12}) yields%
\begin{eqnarray}
G^{Cl}\left( \mathbf{p}_{1},\mathbf{p}_{2},\mathbf{c}_{2}\right)  &=&\tilde{G%
}_{gen}^{Cl}\left( \mathbf{p}_{1},\mathbf{p}_{2},\mathbf{X}\left( \mathbf{p}%
_{1},\mathbf{c}_{2}\right) ,\mathbf{c}_{2}\right)   \notag \\
&=&\mathbf{p}_{1}\cdot \mathbf{X}\left( \mathbf{p}_{1},\mathbf{c}_{2}\right)
+\mathbf{p}_{2}\cdot \mathbf{c}_{2}-F\left( \mathbf{X}\left( \mathbf{p}_{1},%
\mathbf{c}_{2}\right) ,\mathbf{c}_{2}\right) ,  \label{gp12}
\end{eqnarray}%
which can be treated as an explicit form of the Legendre-Clairaut transform
for a non-convex function $F\left( \mathbf{x}\right) $ having degenerate
Hessian matrix. Note that the relation between the Legendre transform and
the parallel curves (which is connected with the above general solution) was
considered a long time ago \cite{ste,har/win}.

Now we apply the Legendre-Clairaut transform to the Hamiltonian procedure
for constrained systems with finite number of degrees of freedom (in the
language of classical mechanics). This is sufficient for exploring the main
idea which can be easily generalized to a field theory (e.g. using DeWitt's
condensed notation \cite{dewitt1}.

\section{The Hamiltonian procedure and the Clairaut equation}

Let $\mathsf{Q}$ be a $n$-dimensional configuration space being a smooth
manifold with local coordinates $\mathbf{q}=\left( q^{1},\ldots q^{n}\right)
$ (all statements can be translated into the coordinate free language \cite%
{car6,tul}). A Lagrangian on $\mathsf{Q}$ is a continuous function $\mathcal{%
L}:T\mathsf{Q}\rightarrow \mathbb{R}$ that is smooth on the tangle bundle $T%
\mathsf{Q}\diagdown \{0\}$ which in local coordinates is determined by%
\footnote{%
As we consider time independent Lagrangians for conciseness, the
time-dependent case can be treated similarly.} $\left( \mathbf{q},\mathbf{v}%
\right) $, where $\mathbf{v}\left( t\right) =\tfrac{d\mathbf{q}\left(
t\right) }{dt}$ are velocities. In this notation the Euler-Lagrange
equations of motion are%
\begin{equation}
\dfrac{d}{dt}\dfrac{\partial \mathcal{L}_{\left( \mathbf{q}\right) }\left(
\mathbf{v}\right) }{\partial \mathbf{v}}-\dfrac{\partial \mathcal{L}_{\left(
\mathbf{q}\right) }\left( \mathbf{v}\right) }{\partial \mathbf{q}}%
=\sum_{i=1}^{n}\left( \mathsf{W}_{\left( \mathbf{q}\right) ij}\left( \mathbf{%
v}\right) \dot{v}^{i}-K_{\left( \mathbf{q}\right) i}\left( \mathbf{v}\right)
\right) =0,  \label{le}
\end{equation}%
where $\mathsf{W}_{\left( \mathbf{q}\right) ij}\left( \mathbf{v}\right)
\overset{def}{=}\partial ^{2}\mathcal{L}_{\left( \mathbf{q}\right) }\left(
\mathbf{v}\right) \diagup \partial v^{i}\partial v^{j}$ is the Hessian
matrix and $K_{\left( \mathbf{q}\right) i}\left( \mathbf{v}\right) \overset{%
def}{=}\tfrac{\partial \mathcal{L}_{\left( \mathbf{q}\right) }\left( \mathbf{%
v}\right) }{\partial q^{i}}-\sum_{j=1}^{n}v^{j}\tfrac{\partial ^{2}\mathcal{L%
}_{\left( \mathbf{q}\right) }\left( \mathbf{v}\right) }{\partial
v^{i}\partial q^{j}}$.

The standard Legendre transformation \cite{arnold} is a local mapping $T%
\mathsf{Q}\rightarrow T^{\ast }\mathsf{Q}$ (the latter is the phase space
which is $\left( \mathbf{q},\mathbf{p}\right) $ in local coordinates) or $%
\mathfrak{Leg}:\mathcal{L}\rightarrow \mathcal{H}$, where $\mathcal{H}%
:T^{\ast }\mathsf{Q}\rightarrow \mathbb{R}$ is a Hamiltonian.

First consider the regular case, when the Hessian $H_{\mathcal{L}%
_{\left( \mathbf{q}\right) }}\left( \mathbf{v}\right) \overset{def}{=}\det
\left\Vert W_{\left( \mathbf{q}\right) ij}\left( \mathbf{v}\right)
\right\Vert $ is nonvanishing. Indeed, the observations of \textbf{Section %
\ref{sec-lt}} imply that it is the Legendre transform in velocities
(considering $\mathbf{q}$ as parameters\footnote{%
We write $\mathbf{q}$-dependence as a subscript to single out coordinates as
passive variables or parameters under the Legendre transformation for which $%
\mathbf{v}$, $\mathbf{p}$ are the active variables.}) such that $\mathcal{H}%
_{\left( \mathbf{q}\right) }\left( \mathbf{p}\right) =\max {}_{\mathbf{v}}%
\mathcal{\tilde{H}}_{\left( \mathbf{q}\right) }\left( \mathbf{p},\mathbf{v}%
\right) $, where $\mathcal{\tilde{H}}_{\left( \mathbf{q}\right) }\left(
\mathbf{p},\mathbf{v}\right) \overset{def}{=}\mathbf{p}\cdot \mathbf{v}-%
\mathcal{L}_{\left( \mathbf{q}\right) }\left( \mathbf{v}\right) $. The
extremum occurs, when%
\begin{equation}
\dfrac{\partial \mathcal{\tilde{H}}_{\left( \mathbf{q}\right) }\left(
\mathbf{p},\mathbf{v}\right) }{\partial \mathbf{v}}=\mathbf{p-}\dfrac{%
\partial \mathcal{L}_{\left( \mathbf{q}\right) }\left( \mathbf{v}\right) }{%
\partial \mathbf{v}}=0,  \label{hl0}
\end{equation}%
which can be resolved with respect to velocities $\mathbf{v}=\mathbf{V}%
_{\left( \mathbf{q}\right) }\left( \mathbf{p}\right) $, since the Hessian is
nonvanishing. Then in the regular case the Hamiltonian is%
\begin{equation}
\mathcal{H}_{\left( \mathbf{q}\right) }\left( \mathbf{p}\right) \overset{def}%
{=}\mathcal{\tilde{H}}_{\left( \mathbf{q}\right) }\left( \mathbf{p},\mathbf{V%
}_{\left( \mathbf{q}\right) }\left( \mathbf{p}\right) \right) =\mathbf{p}%
\cdot \mathbf{V}_{\left( \mathbf{q}\right) }\left( \mathbf{p}\right) -%
\mathcal{L}_{\left( \mathbf{q}\right) }\left( \mathbf{V}_{\left( \mathbf{q}%
\right) }\left( \mathbf{p}\right) \right) .  \label{h}
\end{equation}%
Now we differentiate $\mathcal{H}_{\left( \mathbf{q}\right) }\left( \mathbf{p%
}\right) $ and obtain%
\begin{equation}
\dfrac{\partial \mathcal{H}_{\left( \mathbf{q}\right) }\left( \mathbf{p}%
\right) }{\partial \mathbf{p}}\overset{(\ref{hl0})}{=}\mathbf{V}_{\left(
\mathbf{q}\right) }\left( \mathbf{p}\right) .  \label{hp}
\end{equation}

Because (\ref{hp}) holds for all $\mathbf{q},\mathbf{p}$ identically for a
solution of (\ref{hl0}), we are able to substitute $\mathbf{V}_{\left(
\mathbf{q}\right) }\left( \mathbf{p}\right) $ into (\ref{h}) and obtain the
Clairaut equation for the Hamiltonian as follows (cf. (\ref{g}))%
\begin{equation}
\mathcal{H}_{\left( \mathbf{q}\right) }^{Cl}\left( \mathbf{p}\right) =%
\mathbf{p}\cdot \dfrac{\partial \mathcal{H}_{\left( \mathbf{q}\right)
}^{Cl}\left( \mathbf{p}\right) }{\partial \mathbf{p}}-\mathcal{L}_{\left(
\mathbf{q}\right) }\left( \dfrac{\partial \mathcal{H}_{\left( \mathbf{q}%
\right) }^{Cl}\left( \mathbf{p}\right) }{\partial \mathbf{p}}\right) .
\label{hh}
\end{equation}%
We call this map \textit{a (generalized) Legendre-Clairaut transformation} $%
\mathfrak{Leg}^{Cl}:\mathcal{L}\rightarrow
\mathcal{H}^{Cl}$, because (\ref{hh}) has a
solution also in the case of singular Lagrangians. In the regular case we
follow the steps of the previous section in considering a general solution
of (\ref{hh})%
\begin{equation}
\mathcal{\tilde{H}}_{\left( \mathbf{q}\right) gen}^{Cl}\left( \mathbf{p},%
\mathbf{v}\right) =\mathbf{p}\cdot \mathbf{v}-\mathcal{L}_{\left( \mathbf{q}%
\right) }\left( \mathbf{v}\right) ,  \label{hg}
\end{equation}%
where initially $\mathbf{v=C}_{\left( \mathbf{q}\right) }$ are constants
with respect to the active variables $\mathbf{v}$, $\mathbf{p}$, i.e.
arbitrary functions of $\mathbf{q}$ as parameters. The envelope solution of (%
\ref{hg}) is subject to the extremum condition%
\begin{equation}
\dfrac{\partial \mathcal{\tilde{H}}_{\left( \mathbf{q}\right)
gen}^{Cl}\left( \mathbf{p},\mathbf{v}\right) }{\partial \mathbf{v}}=\mathbf{p%
}-\dfrac{\partial \mathcal{L}_{\left( \mathbf{q}\right) }\left( \mathbf{v}%
\right) }{\partial \mathbf{v}}=0,  \label{hc}
\end{equation}%
which coincides with (\ref{hl0}) and determines additional
dependence on
the momenta when we resolve (\ref{hc}) (which is possible because $H_{%
\mathcal{L}}\neq 0$), and we denote this solution by $\mathbf{v=V}_{\left(
\mathbf{q}\right) }\left( \mathbf{p}\right) $. After substituting into (\ref%
{hg}) we obtain the envelope solution of the Clairaut equation (\ref{hh}) as%
\begin{equation}
\mathcal{H}_{\left( \mathbf{q}\right) env}^{Cl}\left( \mathbf{p}\right)
\overset{def}{=}\mathcal{\tilde{H}}_{\left( \mathbf{q}\right)
gen}^{Cl}\left( \mathbf{p},\mathbf{v}\right) |_{\mathbf{v=V}_{\left( \mathbf{%
q}\right) }\left( \mathbf{p}\right) }=\mathbf{p}\cdot \mathbf{V}_{\left(
\mathbf{q}\right) }\left( \mathbf{p}\right) -\mathcal{L}_{\left( \mathbf{q}%
\right) }\left( \mathbf{V}_{\left( \mathbf{q}\right) }\left( \mathbf{p}%
\right) \right) =\mathcal{H}_{\left( \mathbf{q}\right) }\left( \mathbf{p}%
\right) ,
\end{equation}%
which coincides with the standard Legendre transformation (\ref{h}), as it
should be in the regular case \cite{arnold1}.

\section{The Clairaut equation for constrained systems}

Consider a singular dynamical system for which the Hessian $H_{\mathcal{L}}$
vanishes. Direct application of the standard Legendre transformation is not
possible now, because (\ref{hl0}) cannot be solved with respect to
velocities $\mathbf{v}$. But in the Clairaut equation (\ref{hh}) there are
no restrictions on $\mathcal{L}_{\left( \mathbf{q}\right) }\left( \mathbf{v}%
\right) $ except smoothness, and therefore, as in \textbf{Section \ref%
{sec-lt}}, we are able to consider the (generalized) Legendre-Clairaut
transformation (\ref{hh}) in the singular case $H_{\mathcal{L}_{\left(
\mathbf{q}\right) }}\left( \mathbf{v}\right) =0$ as well.

Let the rank of the Hessian matrix be less than the configuration
space dimension $rank\left\Vert \mathsf{W}_{\left( \mathbf{q}\right)
ij}\left( \mathbf{v}\right) \right\Vert =k<n$ and $k$ is constant. We
rearrange the indices $i$, $j$ in such a way that a non-singular minor of
rank $k$ will be in the upper left-hand corner. Then, we express the index $i
$ as a pair $i=\left( i_{1},i_{2}\right) $, $i_{1}=1,\ldots k$, $%
i_{2}=k+1,\ldots n$, and decompose sets of coordinates and momenta as $%
\mathbf{q}=\left( \mathbf{q}_{1},\mathbf{q}_{2}\right)=\left( \left\{q^{i_{1}}\right\} ,( \left\{q^{i_{2}}\right\}\right) $, $\mathbf{p}=\left(
\mathbf{p}_{1},\mathbf{p}_{2}\right)=\left( \left\{p_{i_{1}}\right\} ,( \left\{p_{i_{2}}\right\}\right)  $, calling the first and the second set
as regular and nonregular coordinates/momenta, respectively. In this
notation the Hessian matrix is $\mathsf{W}_{ij}=\left(
\begin{array}{cc}
\mathsf{W}_{i_{1}i_{1}} & \mathsf{W}_{i_{1}i_{2}} \\
\mathsf{W}_{i_{2}i_{1}} & \mathsf{W}_{i_{2}i_{2}}%
\end{array}%
\right) \overset{def}{=}\left(
\begin{array}{cc}
\mathsf{W}^{\left( 11\right) } & \mathsf{W}^{\left( 12\right) } \\
\mathsf{W}^{\left( 21\right) } & \mathsf{W}^{\left( 22\right) }%
\end{array}%
\right) $, where $\mathsf{W}^{\left( 11\right) }$ is nonsingular $\det
\mathsf{W}^{\left( 11\right) }\neq 0 $, $rank~%
\mathsf{W}^{\left( 11\right) }=k$%
. Then the Clairaut equation (\ref{hh}) acquires the form%
\begin{equation}
\mathcal{H}_{\left( \mathbf{q}\right) }^{Cl}\left( \mathbf{p}\right) =%
\mathbf{p}_{1}\cdot \dfrac{\partial \mathcal{H}_{\left( \mathbf{q}\right)
}^{Cl}\left( \mathbf{p}\right) }{\partial \mathbf{p}_{1}}+\mathbf{p}%
_{2}\cdot \dfrac{\partial \mathcal{H}_{\left( \mathbf{q}\right) }^{Cl}\left(
\mathbf{p}\right) }{\partial \mathbf{p}_{2}}-\mathcal{L}_{\left( \mathbf{q}%
\right) }\left( \dfrac{\partial \mathcal{H}_{\left( \mathbf{q}\right)
}^{Cl}\left( \mathbf{p}\right) }{\partial \mathbf{p}_{1}},\dfrac{\partial
\mathcal{H}_{\left( \mathbf{q}\right) }^{Cl}\left( \mathbf{p}\right) }{%
\partial \mathbf{p}_{2}}\right) ,  \label{hc12}
\end{equation}%
\medskip which we treat as a definition of $\mathcal{H}_{\left( \mathbf{q}%
\right) }^{Cl}\left( \mathbf{p}\right) $ in the case of singular
Lagrangians. We cannot derive this, as in the regular case, because there is
no relation (\ref{hc}) for nonregular variables. A general solution of this
partial differential equation is
\begin{equation}
\mathcal{\tilde{H}}_{\left( \mathbf{q}\right) gen}^{Cl}\left( \mathbf{p},%
\mathbf{v}_{1},\mathbf{v}_{2}\right) =\mathbf{p}_{1}\cdot \mathbf{v}_{1}+%
\mathbf{p}_{2}\cdot \mathbf{v}_{2}-\mathcal{L}_{\left( \mathbf{q}\right)
}\left( \mathbf{v}_{1},\mathbf{v}_{2}\right) ,  \label{hg12}
\end{equation}%
where $\mathbf{v}_{1}=\mathbf{C}_{1\left( \mathbf{q}\right) },\mathbf{v}_{2}=%
\mathbf{C}_{2\left( \mathbf{q}\right) }$ are arbitrary functions of the
passive variables $\mathbf{q}$. As compared to (\ref{hc}), we can find the
envelope solution for the regular part only, i.e. we obtain the
mixed envelope solution. The extremum condition for regular variables now is%
\begin{equation}
\dfrac{\partial \mathcal{\tilde{H}}_{\left( \mathbf{q}\right)
gen}^{Cl}\left( \mathbf{p},\mathbf{v}_{1},\mathbf{v}_{2}\right) }{\partial
\mathbf{v}_{1}}=\mathbf{p}_{1}-\dfrac{\partial \mathcal{L}_{\left( \mathbf{q}%
\right) }\left( \mathbf{v}_{1},\mathbf{v}_{2}\right) }{\partial \mathbf{v}%
_{1}}=0,  \label{pc}
\end{equation}%
which can be solved (due to $\det \mathsf{W}^{\left( 11\right) }\neq 0$) as $%
\mathbf{v}_{1}=\mathbf{V}_{\left( \mathbf{q}\right) }\left( \mathbf{p}_{1},%
\mathbf{v}_{2}\right) |_{\mathbf{v}_{2}=\mathbf{C}_{2}\left( \mathbf{q}%
\right) }=\mathbf{V}_{\left( \mathbf{q}\right) }\left( \mathbf{p}_{1},%
\mathbf{C}_{2\left( \mathbf{q}\right) }\right) $. Then we substitute this
solution into (\ref{hg12}) and obtain the \textquotedblleft
mixed Hamiltonian\textquotedblright\ (or unconstrained Hamiltonian) in the
form%
\begin{eqnarray}
&&\mathcal{H}_{\left( \mathbf{q}\right) mixed}^{Cl}\left( \mathbf{p},\mathbf{v%
}_{2}\right) \overset{def}{=}\mathcal{\tilde{H}}_{\left( \mathbf{q}\right)
gen}^{Cl}\left( \mathbf{p},\mathbf{v}_{1},\mathbf{v}_{2}\right) |_{\mathbf{v}%
_{1}=\mathbf{V}_{\left( \mathbf{q}\right) }\left( \mathbf{p}_{1},\mathbf{v}%
_{2}\right) }  \notag \\
&=&\mathbf{p}_{1}\cdot \mathbf{V}_{\left( \mathbf{q}\right) }\left( \mathbf{p%
}_{1},\mathbf{v}_{2}\right) +\mathbf{p}_{2}\cdot \mathbf{v}_{2}-\mathcal{L}%
_{\left( \mathbf{q}\right) }\left( \mathbf{V}_{\left( \mathbf{q}\right)
}\left( \mathbf{p}_{1},\mathbf{v}_{2}\right) ,\mathbf{v}_{2}\right) ,
\label{hmix}
\end{eqnarray}%
where $\mathbf{v}_{2}=\mathbf{C}_{2\left( \mathbf{q}\right) }$ remain
arbitrary functions of the passive variables $\mathbf{q}$. In this picture
the relation (\ref{pc}) is not a definition of the momenta, but rather a
condition for the existence of the envelope solution for the regular part. A
similar condition for the nonregular part does not exist. Therefore at this
initial stage the nonregular momenta $\mathbf{p}_{2}$ have no connection
with the Lagrangian (analogous to (\ref{pc})), thus now a \textquotedblleft
true\textquotedblright\ (in the standard definition) phase space is formed
by $\left( \mathbf{q}_{1},\mathbf{p}_{1}\right) \in \left( T^{\ast }\mathsf{Q%
}\right) _{1}$ only.

Note that in \cite{tul/urb} the passage from the Lagrangian $\mathcal{L}%
_{\left( \mathbf{q}\right) }\left( \mathbf{v}\right) $ to the general
solution $\mathcal{\tilde{H}}_{\left( \mathbf{q}\right) gen}^{Cl}\left(
\mathbf{p},\mathbf{v}_{1},\mathbf{v}_{2}\right) $ is called
\textquotedblleft a slow and careful Legendre
transformation\textquotedblright , while the further passage to the
mixed Hamiltonian $\mathcal{H}_{\left( \mathbf{q}\right) mixed}^{Cl}\left(
\mathbf{p},\mathbf{v}_{2}\right) $ is called \textquotedblleft a reduction
of the global Hamiltonian Morse family\textquotedblright . Also, the given
Legendre-Clairaut transformation becomes exactly a generalized Legendre
transformation as of \cite{cen/hol/hoj/mar}.

Now we consider the full differential of both sides of (\ref{hmix}) and use
the extremum condition (\ref{pc}), which gives%
\begin{eqnarray}
\left. \dfrac{\partial \mathcal{H}_{\left( \mathbf{q}\right)
mixed}^{Cl}\left( \mathbf{p},\mathbf{v}_{2}\right) }{\partial \mathbf{q}_{1,2}%
}\right\vert _{\mathbf{v}_{2}=\mathbf{C}_{2\left( \mathbf{q}\right) }}
&=&-\left. \dfrac{\partial \mathcal{L}_{\left( \mathbf{q}\right) }\left(
\mathbf{v}_{1},\mathbf{v}_{2}\right) }{\partial \mathbf{q}_{1,2}}\right\vert
_{\substack{ \mathbf{v}_{1}=\mathbf{V}_{\left( \mathbf{q}\right) }\left(
\mathbf{p}_{1},\mathbf{C}_{2\left( \mathbf{q}\right) }\right)  \\ \mathbf{v}%
_{2}=\mathbf{C}_{2\left( \mathbf{q}\right) }}}  \notag \\
&&+\mathbf{R}_{\left( \mathbf{q}\right) }^{\left( 1,2\right) }\left( \mathbf{%
p},\mathbf{C}_{2\left( \mathbf{q}\right) }\right) ,  \label{hm1} \\
\left. \dfrac{\partial \mathcal{H}_{\left( \mathbf{q}\right)
mixed}^{Cl}\left( \mathbf{p},\mathbf{v}_{2}\right) }{\partial \mathbf{p}_{1}}%
\right\vert _{\mathbf{v}_{2}=\mathbf{C}_{2\left( \mathbf{q}\right) }} &=&%
\mathbf{V}_{\left( \mathbf{q}\right) }\left( \mathbf{p}_{1},\mathbf{C}%
_{2\left( \mathbf{q}\right) }\right) ,  \label{hm2} \\
\left. \dfrac{\partial \mathcal{H}_{\left( \mathbf{q}\right)
mixed}^{Cl}\left( \mathbf{p},\mathbf{v}_{2}\right) }{\partial \mathbf{p}_{2}}%
\right\vert _{\mathbf{v}_{2}=\mathbf{C}_{2\left( \mathbf{q}\right) }} &=&%
\mathbf{C}_{2\left( \mathbf{q}\right) },  \label{hm4}
\end{eqnarray}%
where%
\begin{align}
& R_{\left( \mathbf{q}\right)i_{2}}^{\left( 1,2\right) }\left(\mathbf{p},\mathbf{C}%
_{2\left( \mathbf{q}\right) }\right) =\sum_{i_{2}^{\prime }=k+1}^{n}\Phi
_{\left( \mathbf{q}\right) i_{2}^{\prime }}\left( \mathbf{p}\right) \dfrac{%
\partial C_{2\left( \mathbf{q}\right) ,i_{2}^{\prime }}}{\partial
q_{1,2}^{i_{2}}},  \label{r1} \\
& \mathbf{\Phi }_{\left( \mathbf{q}\right) }\left( \mathbf{p}\right) =%
\mathbf{p}_{2}-\mathbf{\Psi }_{\left( \mathbf{q}\right) }\left( \mathbf{p}%
_{1}\right) ,\ \ \mathbf{\Psi }_{\left( \mathbf{q}\right) }\left( \mathbf{p}%
_{1}\right) =\left. \dfrac{\partial \mathcal{L}_{\left( \mathbf{q}\right)
}\left( \mathbf{v}_{1},\mathbf{v}_{2}\right) }{\partial \mathbf{v}_{2}}%
\right\vert _{\substack{ \mathbf{v}_{1}=\mathbf{V_{\left( \mathbf{q}\right)}}\left( \mathbf{}\mathbf{p%
}_{1},\mathbf{C}_{2\left( \mathbf{q}\right) }\right)  \\ \mathbf{v}_{2}=%
\mathbf{C}_{2\left( \mathbf{q}\right) }}}  \label{psi}
\end{align}%
and $\mathbf{C}_{2\left( \mathbf{q}\right) }$ are still arbitrary. Note that
$\mathbf{\Psi }_{\left( \mathbf{q}\right) }\left( \mathbf{p}_{1}\right) $
and therefore $\mathbf{\Phi }_{\left( \mathbf{q}\right) }\left( \mathbf{p}%
\right) $ have no dependence on the unsolved velocities $\mathbf{v}_{2}$,
because, if some of them appeared there, we could derive them from (\ref{psi}%
), which contradicts the fact that the rank of the Hessian is $k$. Then, using the Lagrange
equations (\ref{le}) and $\mathbf{\dot{q}}_{1}=\mathbf{V_{\left( \mathbf{q}\right)}}\left( \mathbf{}\mathbf{p%
}_{1},\mathbf{C}_{2\left( \mathbf{q}\right) }\right)$, 
$\mathbf{\dot{q}}_{2}=\mathbf{C}_{2\left( \mathbf{q}\right)}$, we obtain the \textquotedblleft mixed Hamiltonian equations of
motion\textquotedblright\ (or unconstrained Hamiltonian equations)%
\begin{eqnarray}
\dfrac{\partial \mathcal{H}_{\left( \mathbf{q}\right) mixed}\left( \mathbf{p}\right) }{\partial \mathbf{q}_{1}} &=&-\mathbf{\dot{p}}%
_{1}+\mathbf{R}_{\left( \mathbf{q}\right) }^{\left( 1\right) }\left( \mathbf{%
p},\mathbf{\dot{q}}_{2}\right) ,  \label{hs1} \\
\dfrac{\partial \mathcal{H}_{\left( \mathbf{q}\right) mixed}\left( \mathbf{p}\right) }{\partial \mathbf{p}_{1}} &=&\mathbf{\dot{q}}_{1},  \label{hs2} \\
\dfrac{\partial \mathcal{H}_{\left( \mathbf{q}\right) mixed}\left( \mathbf{p}\right) }{\partial \mathbf{q}_{2}} &=&-\mathbf{\dot{\Psi }}_{\left( \mathbf{q}\right) }\left( \mathbf{p}%
_{1}\right) +\mathbf{R}_{\left( \mathbf{q}\right) }^{\left( 2\right) }\left( \mathbf{p},%
\mathbf{\dot{q}}_{2}\right) ,  \label{hs3} \\
\dfrac{\partial \mathcal{H}_{\left( \mathbf{q}\right) mixed}\left( \mathbf{p}\right) }{\partial \mathbf{p}_{2}} &=&\mathbf{\dot{q}}%
_{2},  \label{hs4}
\end{eqnarray}%
It can be shown that the system (\ref{hs1})--(\ref{hs4}) leads to the
equations of motion which are equivalent to the Lagrangian ones (\ref{le}).
Observe that, if we use (\ref{pc}) also for nonregular velocities and momenta,
we can  present r.h.s. of the equation (\ref{hs3}) as $-\mathbf{\dot{p}}%
_{2}+\mathbf{\dot{\Phi}}_{\left( \mathbf{q}\right) }\left( \mathbf{p}\right)
+\mathbf{R}_{\left( \mathbf{q}\right) }^{\left( 2\right) }\left( \mathbf{p},%
\mathbf{\dot{q}}_{2}\right) $. Then (\ref{hs1})--(\ref{hs4}) becomes the standard system of Hamiltonian
equations, when the following system of equations (generalized constraints)
is valid%
\begin{equation}
\mathbf{R}_{\left( \mathbf{q}\right) }^{\left( 1\right) }\left( \mathbf{p},%
\mathbf{\dot{q}}_{2}\right) =0,\ \ \ \ \ \ \mathbf{\dot{\Phi}}_{\left(
\mathbf{q}\right) }\left( \mathbf{p}\right) +\mathbf{R}_{\left( \mathbf{q}%
\right) }^{\left( 2\right) }\left( \mathbf{p},\mathbf{\dot{q}}_{2}\right) =0
\label{fr}
\end{equation}%
The generalized constraints (\ref{fr}) are sufficient to introduce the
standard Poisson brackets and the \textquotedblleft
correct\textquotedblright\ time evolution \cite{arnold}.

In the Dirac formalism \cite{dirac} one imposes%
\begin{equation}
\mathbf{\Phi }_{\left( \mathbf{q}\right) }\left( \mathbf{p}\right) =0,\ \ \
\ \ \ \ \ \mathbf{\dot{\Phi}}_{\left( \mathbf{q}\right) }\left( \mathbf{p}%
\right) =0,  \label{ff}
\end{equation}%
which are the standard primary scleronomous constraints (which should be
functionally independent, otherwise see \cite{mik/zan}). Obviously then, the
conditions in (\ref{ff}) are more restrictive than those in (\ref{fr}). But,
they lead to the \textquotedblleft correct\textquotedblright\ phase space in
regular and nonregular variables, in which for both sets of momenta the
derivatives of the Lagrangian with respect to corresponding velocities, are
initially treated as definitions \cite{car6}. Indeed, this allows one to
consider $\left( \mathbf{q},\mathbf{p}\right) $ as points of the entire
\textquotedblleft true\textquotedblright\ phase space $T^{\ast }\mathsf{Q}$,
while the Legendre transformation becomes then a degenerate mapping with a
kernel \cite{car6}. Then the \textquotedblleft
mixed Hamiltonian\textquotedblright\ (\ref{hmix}) can be presented as the
sum of $\mathcal{H}_{\left( \mathbf{q}\right) }^{\left( 0\right) }\left(
\mathbf{p}_{1}\right) $ and the linear combination of the primary constraints%
\begin{align}
& \partial \mathcal{H}_{\left( \mathbf{q}\right) mixed}^{Cl}\left( \mathbf{p},%
\mathbf{C}_{2\left( \mathbf{q}\right) }\right) =\mathcal{H}_{\left( \mathbf{q%
}\right) }^{\left( 0\right) }\left( \mathbf{p}_{1}\right) +\mathbf{C}%
_{2\left( \mathbf{q}\right) }\cdot \mathbf{\Phi }_{\left( \mathbf{q}\right)
}\left( \mathbf{p}\right) ,  \label{hmd} \\
& \mathcal{H}_{\left( \mathbf{q}\right) }^{\left( 0\right) }\left( \mathbf{p}%
_{1}\right) =\mathbf{p}_{1}\cdot \mathbf{V}_{\left( \mathbf{q}\right)
}\left( \mathbf{p}_{1},\mathbf{C}_{2\left( \mathbf{q}\right) }\right) -%
\mathcal{L}_{\left( \mathbf{q}\right) }\left( \mathbf{V}_{\left( \mathbf{q}%
\right) }\left( \mathbf{p}_{1},\mathbf{C}_{2\left( \mathbf{q}\right)
}\right) ,\mathbf{C}_{2\left( \mathbf{q}\right) }\right) \nonumber \\
& +\mathbf{C}_{2\left( \mathbf{q}\right) }\cdot \mathbf{\Psi }_{\left(
\mathbf{q}\right) }\left( \mathbf{p}_{1}\right) ,
\end{align}%
where $\mathcal{H}_{\left( \mathbf{q}\right) }^{\left( 0\right) }\left(
\mathbf{p}_{1}\right) $ does not depend on $\mathbf{p}_{2}$ and $\mathbf{C}%
_{2\left( \mathbf{q}\right) }$ due to (\ref{pc}) and (\ref{hm4}). In this
case (\ref{hmd}) coincides with Dirac's total Hamiltonian  \cite{dirac}.

If one does not impose the generalized constraints (\ref{fr}), then the
equation (\ref{hs3}) has no time derivative of the nonregular momenta $%
\mathbf{p}_{2}$ on the right-hand side. Thus (\ref{hs3}) is an algebraic
equation for $\mathbf{p}_{2}$, and therefore there are no standard Hamiltonian
equations for nonregular variables. Nevertheless, the system of the
\textquotedblleft mixed Hamiltonian equations of motion\textquotedblright\
describes the same constrained physical system, as Lagrangian equations of
motion (\ref{le}), independently of whether or not the constraints (\ref{ff}%
) are imposed.

Finally, we can show that the Legendre-Clairaut transformation is involutive
for constrained systems. Indeed, in full analogy with (\ref{hmix}), for a
given \textquotedblleft mixed Hamiltonian\textquotedblright , satisfying (\ref%
{hs1})--(\ref{hs4}), we can construct the corresponding Clairaut equation
(similar to (\ref{hc12})), using the fact that the Hessian matrices for $%
\mathcal{L}$ and $\mathcal{H}$ have the same rank (see \textbf{Section \ref%
{sec-lt}}), from which we obtain its mixed envelope solution. This can then
be called a \textquotedblleft mixed Lagrangian\textquotedblright\ (or
unconstrained Lagrangian)%
\begin{equation}
\mathcal{L}_{\left( \mathbf{q}\right) mixed}^{Cl}\left( \mathbf{v},%
\mathbf{p}_{2}\right) =\mathbf{v}_{1}\cdot \mathbf{P}_{\left( \mathbf{q}%
\right) }\left( \mathbf{v}_{1},\mathbf{p}_{2}\right) +\mathbf{v}_{2}\cdot
\mathbf{p}_{2}-\mathcal{H}_{\left( \mathbf{q}\right) mixed}^{Cl}\left( \mathbf{P}%
_{\left( \mathbf{q}\right) }\left( \mathbf{v}_{1},\mathbf{p}_{2}\right) ,%
\mathbf{p}_{2},\mathbf{v}_2\right) ,
\end{equation}%
where $\mathbf{P}_{\left( \mathbf{q}\right) }\left( \mathbf{v}_{1},\mathbf{p}%
_{2}\right) =\mathbf{p}_{1}$ is a solution of the  condition $%
\mathbf{v}_{1}=\tfrac{\partial \mathcal{H}_{\left( \mathbf{q}\right)
mixed}^{Cl}\left( \mathbf{p},\mathbf{v}_{2}\right) }{\partial \mathbf{p}_{1}}$
similar to (\ref{p}), and $\mathbf{p}_{2}$ are arbitrary functions (unsolved
momenta). If the functions $\mathbf{P}_{\left( \mathbf{q}\right) }\left(
\mathbf{v}_{1},\mathbf{p}_{2}\right) $ and $\mathbf{V}_{\left( \mathbf{q}%
\right) }\left( \mathbf{p}_{1},\mathbf{v}_{2}\right) $ are mutually inverse
in regular variables, i.e. $\mathbf{P}_{\left( \mathbf{q}\right) }\left(
\mathbf{V}_{\left( \mathbf{q}\right) }\left( \mathbf{p}_{1},\mathbf{v}%
_{2}\right) ,\mathbf{p}_{2}\right) =\mathbf{p}_{1}$, $\mathbf{V}_{\left(
\mathbf{q}\right) }\left( \mathbf{P}_{\left( \mathbf{q}\right) }\left(
\mathbf{v}_{1},\mathbf{p}_{2}\right) ,\mathbf{v}_{2}\right) =\mathbf{v}_{1}$%
, then it can be verified that $\mathcal{L}_{
mixed}^{Cl}=\mathcal{L} $, which proves involutivity.

\section{Conclusion}

To summarize: the concise treatment of constrained systems using the Clairaut
partial differential equation presented here gives a different  explanation
of the Dirac primary constraints from the Lagrange multiplier method, as well as a
different understanding of the nature of primary constraints. In some cases
this can lead to possible generalizations. Moreover, it can be applied in
the cases where the standard constraint methods do not work or are too
cumbersome. Examples, details and corresponding Hamiltonian formalism 
will appear in the forthcoming paper.

\textit{Acknowledgements}. The author would like to express his thanks to
V.~P.~Aku\-lov, A.~V.~Antonyuk, Yu.~A.~Berezhnoj, A.~T.~Kotvitskij,
G.~C.~Kurinnoj, P.~Ma\-hnke, B.~V.~Novikov, L.~A.~Pastur, S.~A.~Ovsienko,
S.~V.~Peletminskij, S.~V.~Prokushkin, A.~S.~Sadovnikov,
S.~D.~Sinel'\-shchikov, W.~Siegel, K.~S.~Stelle, R.~Wul\-ken\-haar and
A.~A.~Zheltukhin for their fruitful discussions. He is also grateful to the
Alexander von Humboldt Foundation for its valuable support and to M.
Zirnbauer for his kind hospitality at the Institute of Theoretical Physics,
Cologne University, where this paper was completed.

\end{document}